\documentstyle[epsf,preprint,aps]{revtex}
\tightenlines
\begin{document}
\draft

\title{Neutrino Propagation and Spin Zero Sound 
in Hot Neutron Matter with Skyrme Interactions}
\author{J. \ Navarro}
\address{IFIC (CSIC - Universidad de Valencia), Facultad de
F{\' \i}sica, 46100 Burjassot, Spain}
\author{E. S. \ Hern\'andez}
\address{Departamento de F{\' \i}sica, Universidad de Buenos Aires,
1428 Buenos Aires, Argentina}
\author{and  \\  D. \ Vautherin}
\address{LPTPE, Universit\'e P. \& M. Curie, case 127, 4 Place Jussieu,
75252 Paris Cedex 05, France}
\date{\today}
\maketitle

\begin{abstract}
We present microscopic calculations of neutrino propagation in
hot neutron matter above nuclear density within the framework of
the Random Phase Approximation. Calculations are performed  for
non-degenerate neutrinos using various Skyrme effective interactions.
We find that for densities just above nuclear density,
spin zero sound is present at zero temperature 
for all Skyrme forces considered. However it disappears rapidly with
increasing temperature due to a strong Landau damping. As a
result the mean-free path is given, to a good approximation, by the mean
field value.
Because of the renormalization of the bare 
mass in the mean field, the medium is more transparent 
as compared to the free case. 
We find, in contrast, that at several times nuclear density, 
a new type of behavior sets in due to the vicinity of a magnetic instability.
It produces a strong reduction of the mean free path.
The corresponding transition density however occurs in a region where inputs
from more realistic calculations are necessary for the construction of a
reliable Skyrme type parametrization. 
\end{abstract}




\section{Introduction}
The gravitational collapse of massive stars at the end of their
thermonuclear burning is believed to produce a core of hot dense
matter in which most of the initial gravitational binding energy
is stored into neutrinos.  Numerical simulations of the
subsequent evolution, leading to the formation of a neutron
star, require the knowledge of the equation of state of hot
dense matter as well as a reliable description of neutrino
transport phenomena. Relevant quantities are 
the specific
heat of the medium and the neutrino mean free path.

The mean free path of a neutrino due to scattering inside
neutron matter at temperature T is proportional to the optical
potential. It can be expressed in the case of non-degenerate
neutrinos as \cite{iwamoto}
\begin{equation}
\frac{1}{\lambda({\bf k}_i, T)} = 
\frac{G_F^2}{32 \pi^3 (\hbar c)^4} \int  d {\bf k}_f 
\left[ (1+\cos \theta) S^{(0)}(\omega, {\bf q}, T)  
+ g_A^2 (3-\cos\theta) \, S^{(1)}(\omega,{\bf q},T) \right]
\label{opa}
\end{equation}
where $G_F$ is the Fermi constant, $g_A$ the axial coupling
constant, ${\bf k}_i$ and ${\bf k}_f$ are the initial and final
neutrino momenta, ${\bf q}$ is the transferred momentum ${\bf
k}_i-{\bf k}_f$, $\omega$ is the transferred energy $|{\bf k}_i|
-|{\bf k}_f|$, and $\cos\theta={\hat{\bf k}}_i\cdot{\hat{\bf
k}}_f$.  The functions $S^{(S)}(\omega,{\bf q},T)$   represent
the dynamical structure factors in the spin symmetric ($S$=0) or
spin antisymmetric ($S$=1) channels. They are given by the
imaginary part of the respective response function
$\chi^{(S)}(\omega,{\bf q})$ and contain the relevant
information on the medium.  The vector (axial) part of the
neutral current gives rise to density (spin-density)
fluctuations, corresponding to the $S$=0 ($S$=1) channel.
Although we shall consider only nondegenerate neutrinos, we note
that the previous formula should be corrected for additional
Pauli blocking factors in the degenerate case.

Several authors have evaluated the neutrino mean free path using
various approximation schemes and various models of the trapping
environment. In their early calculations Tubbs and Schramm
\cite{tubbs} omitted the effect of nucleon-nucleon interactions,
whose importance was later on emphasized by Sawyer \cite{sawyer}
showing that for standard equations of state, the neutrino
opacity due to  coherent scattering off density fluctuations can
be significantly larger than in a neutron Fermi gas. Iwamoto
and Pethick\cite{iwamoto} investigated the effect of
nucleon-nucleon interaction within the framework of Landau
theory of Fermi liquids, considering both density and
spin-density fluctuations according to Eq. (\ref{opa}).  The
effective interaction was taken into account by  the monopolar
Landau parameters $F_0$ and $G_0$ derived earlier by B\"ackman
et al.\cite{backman}.  The main conclusion of their work is that
for degenerate neutrinos, there is a reduction of in-medium
scattering cross sections by a factor of 2-3 due to
nucleon-nucleon interactions, and a subsequent increase of the
neutrino mean free path.  Calculations have also been performed
using the ring approximation \cite{pons,BURROWS}, in which the
response function $\chi^{(S)}(\omega,{\bf q})$ of the interacting
system is expressed in terms of the bare response function $\chi
_0(\omega,{\bf q})$ as
\begin{equation} \label{ring}
\chi^{(S)}(\omega,{\bf q})= \frac{\chi_0(\omega,{\bf q})}
{1-V^{(S)}({\bf q})\chi_0(\omega,{\bf q})}, 
\end{equation}
where $V^{(S)}({\bf q})$ is the interaction in momentum space in
spin channel $S$.

The case of non-degenerate neutrinos was examined  by Haensel
and Jerzak \cite{haensel}, who computed both the scattering and
the absorption rates within the framework of Landau theory.
They used a quasiparticle interaction derived from the Reid soft
core nucleon-nucleon potential.  Their results indicate that the
higher the density of the medium, the larger  the effect of the
interaction. More recently, Reddy {\it et al} \cite{pons}
considered degenerate neutrinos immersed in an environment of
nucleons interacting via a Skyrme two-body interaction out of
which the monopole Landau parameters $F_0$ and $G_0$ are
constructed.  These authors consider the Skyrme parametrizations
SGII and SkM$^*$ and simulate the momentum dependence of the
force through the direct matrix element  of the meson exchange
potential.

In this paper we extend preliminary results presented in Ref.
\cite{INPC98} and report systematic, fully self-consistent
calculations based on effective interactions of the Skyrme type,
including some of the most recent ones.  Such interactions have
been very successful at describing properties of finite nuclei
and nuclear as well as neutron matter.  An attractive feature of
Skyrme interactions is that response functions can be calculated
in closed form \cite{garcia,braghin,HNPV} in the Random Phase
Approximation (RPA), which generalizes the simple ring
approximation (\ref{ring}) based on Hartree-like dynamics, to
the more realistic case of Hartree-Fock mean field dynamics.  An
advantage of such a framework is that it also avoids splitting
structure factors into single particle and collective
contributions based on different approximation schemes.  It
further avoids postulating a factorization of the temperature
dependence of the structure factors, which  are in conflict with
the fact that zero sound disappears at high temperature
\cite{hnp}.  Moreover, sum rules hold in the case of fully self-
consistent calculations, thus allowing the possibility of
consistency checks.

In the following we focus on the evaluation of the mean free
path of neutrinos interacting with neutrons through neutral
currents only.  This coupling is important mainly after
deleptonization has taken place and it affects the late
behaviour of the neutrino burst, where neutrino diffusion is
driven by the temperature gradient. Indeed, in the earlier
stages, just after the formation of the protoneutron star,
charged current reactions leading to neutrino absorption are
dominant, due to the fact that  the high temperatures involved
prevent efficient inhibition of charged reactions by the
energy-momentum conservation constraints.  We consider  neutron
matter above nuclear density $\rho_0$=0.17 nucleons/fm$^3$. The
case of high densities is of special interest because medium
effects are expected to play an important role. In contrast,
below nuclear density propagation is dominated by coherent
scattering off nuclei, a case for which reliable estimates of
the opacities are available.

This paper is organized as follows. In Sec. II we briefly
summarize the formalism for the computation of the response of
the neutron liquid and analyze the main results in Sec. III. Our
conclusions and perspectives are presented in Sec. IV.

\section{Response function of hot neutron matter}

In the present work we use the techniques developed in
Refs.\cite{braghin,HNPV} to obtain RPA susceptibilities at
finite temperature, in the particular case of Skyrme-type
effective interactions.  The RPA dynamical susceptibility
$\chi^{(S)}(q,\omega,T)$ in spin channel $S$ of a Fermi system
in thermal equilibrium at temperature $T$ is given by the
the polarizability to an infinitesimal external field
\begin{equation} \label{external}
{\hat V}_{\rm ext}= \varepsilon {\hat {\bf O}} \exp (i \omega t - i {\bf q}.{\bf x}).
\end{equation}
In this equation $\omega$ contains a small imaginary part to ensure an 
adiabatic switching of the external field starting 
at time $t= -\infty$ when the medium is in its equilibrium state at
temperature $T$.
The operator 
${\hat {\bf O}}$ is equal to the unit matrix in spin space for
spin zero (S=0) and to the third Pauli matrix $\sigma_z$ for S=1. 
At late enough times (and small enough $\varepsilon$) the expectation value of
the operator ${\hat {\bf O}}$ has the same space and time dependence as the external
field. The dynamical susceptibility is defined by
\begin{equation} \label{polar}
\chi^{(S)} = \lim_{\varepsilon \to 0} 
\frac{\langle {\hat {\bf O}} \rangle}
{ \varepsilon \exp (i \omega t - i {\bf q}.{\bf x})} , 
\end{equation}
and it can be shown to be given by the 
algebraic expression
\begin{equation}\label{chieq}
\chi^{(S)} = \chi_0
+ G_0 \tilde{V}^{(S)}_{ph} G_0 , 
\end{equation}
where $\chi_0$ is the response of the free system, $G_0$ is the
free particle-hole (ph) propagator at temperature $T$ and
$\tilde{V}^{(S)}_{ph}$ is the RPA induced interaction, related
to the effective ph one $V_{ph}^{(S)}$ by the integral equation
\begin{equation}\label{induph}
\tilde{V}^{(S)}_{ph} = V^{(S)}_{ph} +
V^{(S)}_{ph}\, G_0\,\tilde{V}^{(S)}_{ph}.
\end{equation}
In Ref. \cite{garcia},  it has been shown that for the general
class of ph interactions of the following form containing a
monopole and a dipole term
\begin{equation}\label{vph}
V^{(S)}_{ph}(q_1, q_2; q) = W_1^{(S)}(q) + W_2^{(S)}(q) ({\bf
q}_1 - {\bf q}_2)^2 \,\, ,
\end{equation}
to which the Skyrme ones belong, the solution of the integral
equation  can be algebraically obtained in a closed form, in
terms of generalized Lindhard functions. This result was derived
for the zero temperature case.  However the generalization to a
thermally excited liquid is straightforward.  The effect of
temperature appears implicitly in the RPA equations  through the
Fermi-Dirac occupation numbers contained in the free ph
propagator.  The algebraic structure of both the RPA equation
and its solution is preserved\cite{braghin,HNPV} and the
dynamical susceptibility in channel $(S)$ has the form
\cite{garcia}
\begin{equation}\label{chi}
\chi^{(S)}(q,\omega,T) =
2\,{\frac{\chi_0(q,\omega,T)}{D^{(S)}(q,\omega,T)}},
\end{equation}
where the factor 2 takes into account  the spin degeneracy and
\begin{eqnarray}
D^{(S)}(q , \omega,T)  &=&  1 - W_1^{(S)}(q)\,\chi_0
- 2 W_2^{(S)}  \, \left[{\frac{\displaystyle q^2}{\displaystyle
4}}- {\frac
{\displaystyle
\omega m^*}{\displaystyle q}}^2\, {\frac{\displaystyle
1}{\displaystyle 1- {\frac{\displaystyle
m^{*} k_F^3}{\displaystyle 3
\pi^2}} W_2^{(S)}  }}\right]\,\chi_0
\nonumber \\
&& +  2 W_2^{(S)}\,\left(\frac {q^2}{2}
\,\chi_0 - k_F^2 \, \chi_2\right) \nonumber \\
&& + [W_2^{(S)}\,k_F^2]^2
 \left[
\chi_2^2 - \chi_0\,\chi_4
+ \left({\frac {\omega m^*}{k_F^2}}\right)^2 \,\chi_0^2 - \frac{
m^{*} }{6
\pi^2 k_F}\,q^2\,\chi_0 \right]. 
\end{eqnarray}
In this expression $k_F$ is the neutron Fermi momentum while 
$m^*$ denotes the nucleon effective mass in
the mean field.

Since in pure neutron matter the isospin exchange operator
$P_{\tau}$ reduces identically to the unit matrix,
only particular
combinations of the Skyrme interaction parameters $t_i$ and
$x_i$, namely
\begin{eqnarray}
s_0&=& t_0(1-x_0),~~~~~s_1= t_1(1-x_1) \nonumber \\
s_2&=& t_2(1+x_2),~~~~~s_3= t_3(1-x_3) , 
\end{eqnarray}
are required. Thus, one obtains the following expressions for 
the neutron effective mass 
\begin{equation} \label{EFFMASS}
\frac{\hbar^2}{2m^*}= \frac{\hbar^2}{2m} + \frac{1}{8}
(s_1+3 s_2 ) \rho, 
\end{equation}
where $\rho$ is the density of neutron matter, and for the 
coefficients $W_{1,2}^{(S)}$ of the effective ph interaction
\begin{eqnarray}
W_1^{(0)}&=& s_0 + \frac{(\gamma
+1)(\gamma+2)}{12} s_3
\rho^{\gamma} + \frac{1}{4} [ s_1-3s_2]
q^2 \nonumber \\
W_2^{(0)}&=& \frac{1}{4} [ s_1 + 3s_2] \nonumber \\
W_1^{(1)}&=& -s_0 - \frac{1}{6} s_3
\rho^{\gamma} - \frac{1}{4} [ s_1 +s_2]
q^2\nonumber \\
W_2^{(1)}&=& \frac{1}{4} [ - s_1 + s_2] , 
\end{eqnarray}
with $\gamma$ being 
the exponent of the density dependent term in the Skyrme force
\begin{equation} \label{DDTERM}
\frac{t_3}{6} \rho^{\gamma}({\bf r}_{12}) 
\,(1 + x_3\,P^{\sigma}_{12})\,\delta({\bf r}_{12}).
\end{equation}
In the special case of the Skyrme interaction SIII we have
considered it as a three-body force rather than a density
dependent one, i.e. we have used a value $s_3=0$ in the previous
equations.

The functions $\chi_{2i}(\omega, q)$ ($i$=0, 1, 2) are generalized
susceptibilities per unit volume $\Omega$, defined as \cite{garcia}
\begin{eqnarray}
\chi_{2i}& = &{\frac{1}{\Omega}}\,\sum_{k}
{\frac{1}{2}} \left[ \left({\frac{{\bf k}^{\,
2}}{k_F^{2}}} \right)^{i}+ \left({\frac{({\bf k}+{\bf q})^{2}}
{k_F^{2}}} \right)^{i}\right]
\left(1-n({\bf k}+{\bf q})\right) \,
\nonumber \\
& & n({\bf k}) \left( {\frac{1}{\omega-\omega({\bf k},{\bf q})+i\eta}} -
{\frac{1}{\omega +\omega({\bf k},{\bf q})+i\eta}} \right),
\end{eqnarray}
with 
\begin{equation} \label{FERMI}
n({\bf k}) = \frac{1}{1+\exp(\varepsilon({\bf k}) - \mu)/T} 
\end{equation}
the
Fermi-Dirac occupation probability, $\mu$ being the chemical
potential at the given temperature.  Here
\begin{equation} \label{OMEGA}
\omega({{\bf k},{\bf q}})= 
\varepsilon({{\bf k}+{\bf q}})- \varepsilon({\bf k }),
\end{equation}
with 
\begin{equation} \label{SPENERGY}
\varepsilon({\bf k})$= $ \hbar^2 k^2/2m^*,
\end{equation} 
the single-particle energy.  The dynamical structure factor is
obtained, at both positive and negative energies $\omega$, from
the detailed balance relationship
\begin{equation}
S^{(S)}(q,\omega,T) = - {\frac{1}{\pi}}{\frac{
{\Im}m \, \chi^{(S)}(q,\omega,T)}{1 - e^{-\omega/T}}}.
\end{equation}
It is worthwhile stressing that opposite to the ring
approximation defined by equation (\ref{ring}), which involves
only the Linhard function $\chi_0$, new terms involving the
generalized susceptibility $\chi_2$ and $\chi_4$ arise in the
full RPA approximation.

Another important point about our self- consistent approach is 
that it preserves sum rules. In particular one has the 
following energy weighted sum rule in the spin one case
\begin{equation} \label{sumrule}
\int_{-\infty}^{+\infty} S^{(S=1)}(q,\omega,T) \omega d \omega = 
\{\frac{\hbar^2}{2m^*}+\frac{1}{8} (s_1-s_2) \rho \} \rho q^2.
\end{equation}
In the spin zero case the right hand side is simply 
$\hbar^2 q^2 \rho/ 2m$.

In the following discussions regarding the appearance of zero 
sound modes it will turn out to be convenient to have the 
explicit expressions of the real and imaginary parts of the 
bare response function. Its imaginary part is given by
\begin{equation} \label{IMCHI}
\Im m \chi_{0}(\omega,q) = - \frac{m^2}{2 \pi q} 
\frac{kT}{1 -e^{- \beta \omega}} \log 
\frac{1+e^{\beta (A+ \omega/2)} }{1+ e^{\beta(A- \omega/2)}} .
\end{equation}
In this equation $\beta$= 1/$kT$ and the quantity $A$ is given by
\begin{equation} \label{97e5}
A = \mu - \frac{ m \omega^2}{2q^2} - \frac{q^2}{8 m^2}.
\end{equation}
The expression of the real part is a little more difficult 
to construct. In the zero temperature case it reads \cite{walecka}
\begin{equation} \label{RECHI}
\Re e \chi_{0}(\omega,q, k_F,T=0)= \frac{1}{4 \pi^2} \frac{mk_F^2}{q\hbar^2}
\{-\frac{2q}{k_F}+ \varphi(x_+)-\varphi(x_-) \}.
\end{equation}
In this equation the quantities $x_\pm$ are defined by
\begin{equation} \label{97e7}
x_\pm= \frac{m \omega}{\hbar k_F q} \pm \frac{q}{2k_F},
\end{equation}
and the function $\varphi$ reads
\begin{equation} \label{97e8}
\varphi(x)=(1-x^2) \log|\frac{x-1}{x+1}|.
\end{equation}
For a non zero temperature the real part of the response 
function is obtained by performing an average of the zero 
temperature results obtained for various values $k$ of the 
Fermi momentum with a weight factor equal to the occupation 
number $n(k)$:  
\begin{equation} \label{97e9}
\Re e \chi_{0}(\omega,q, k_F,T)=-\int_0^\infty
\Re e \chi_{0}(\omega,q, k_F=k,T=0) dn(k).
\end{equation}

\section{Results}

We have considered two standard Skyrme parametrizations, namely
SIII \cite{sk3} and SkM$^*$ \cite{skm}, as well as
parametrizations SLy230a and SLy230b recently adjusted
\cite{slb} to reproduce the latest equation of state for pure
neutron matter calculated by Wiringa {\it et al} \cite{wiri}
within the Fermi Hypernetted Chain scheme, using the Urbana V14
two-body interaction plus the Urbana three-body interaction.
The energies per particle given by parametrizations SLy230a,b and by
the realistic force are very close to each other up to densities
of about 1 nucleon/fm$^3$.  In this section we analyze in detail
the evolution of the mean free path with neutron density, its
dependence upon temperature and neutrino momentum and the effect
of residual interactions, as follows.
 
\subsection{Evolution of the mean-free path with density}

In Fig. 1 we show the evolution of the mean free path with
density for an incoming neutrino momentum $k_i$= 5 MeV/c and a
temperature $T$=5 MeV. Results are displayed for the four
effective interactions mentioned above. It can be noticed that
for interactions SLy230a and SLy230b the mean free path exhibits
a slow variation with density. This behavior is similar to the
one obtained by Haensel and Jerzak \cite{haensel} who find mean
free paths $\lambda$= 2, 1.5 and 1km for densities 0.2, 0.36 and
0.6 fm$^{-3}$ respectively, for $k_i c=T=5$MeV.  In contrast the
results obtained with interactions SIII and SkM$^*$ show a
substantially different behavior, both exhibiting a rapid
monotonic decrease with density.  This decay is the consequence
of a pronounced increase in the magnetic strength, that reflects
in the vicinity of an instability in the channel S=1. The magnetic
instability can be located by examining the Landau parameters of
neutron matter, in particular, the Landau coefficient $G_0$
related to the spin asymmetry coefficient $a_{\sigma}$ in
neutron matter through the relation
\begin{equation} \label{LANDAU}
a_{\sigma}= \frac{\hbar^2 k_F^2}{6m^*}(1+G_0).
\end{equation}

In terms of the force parameters, the asymmetry coefficient is
given by the following expression
\begin{equation} \label{ASYM}
a_{\sigma}= \frac{\hbar^2 k_F^2}{6m^*} - 
(s_0+\frac{s_3}{6} \rho^{\gamma}) \frac{\rho}{4}+
\frac{5}{24} (s_2-s_1) \tau.
\end{equation}
where $\tau=3/5\,\rho\,k_F^2$ is the kinetic energy density. For
interaction SIII the last term  dominates at high values of the
neutron density and as a result the Landau parameter takes the
value --1 at a critical density around 0.317 fm$^{-3}$. This
critical density $\rho_c$ is (in fm$^{-3}$) 0.1967 for SkM$^*$,
0.5411 for SLy230a and 0.5928 for SLy230b. These are precisely
the abscissae at which the neutrino mean free path drops to a
very small value in Fig. 1.  Actually, the mean-free path at the
minimum is  not exactly zero, since  even though the monopole
contribution vanishes there remains a small contribution from
the dipole term.

\subsection{Spin zero-sound with Skyrme forces}

To illustrate this point, in Fig. 2 we display the dynamical
structure factors computed with interaction SIII, for spin
channels S=0 and 1 (left and right parts of the figure,
respectively), as functions of energy $\omega$, for transferred
momenta q equal to 5 and 100 MeV/c (dashed and solid lines,
respectively) and a temperature of 5 MeV.  The upper and lower
parts in this figure correspond to  densities 0.1 fm$^{-3}$ and
$\rho_c$.  The zero sound mode is visible in the S=1, $\rho$= 0.1 fm$^{-3}$ 
case although it is smeared by Landau damping at a temperature 
$T$=5 MeV. We also clearly
appreciate the significant increase and change of scale in
S$^{(1)}( \omega, q)$ as the density reaches the transition
point, which contrasts with the rather moderate variation undergone
by S$^{(0)}$ with density.  The numerical data indicate a scale
factor of three orders of magnitude in S$^{(1)}$ as the density
reaches $\rho_c$. 
For S=0 one also observes a peak which corresponds to the maximum in the
imaginary part of the bare response. This peak is reduced 
and shifted towards lower energies in the RPA response
because of the contribution of the real part.

In order to have a qualitative criterion for the existence of
a zero sound mode it is convenient to calculate the response in
the monopole approximation.  In this approximation one evaluates the particle
-hole interaction strength at the Fermi surface and one ignores its angular
dependence i.e. one sets $q_1^2$= $q_2^2$= $k_F^2$ and ${\bf q_1}.{\bf q_2}$=0
in Eq. (\ref{vph}). One thus uses Eq. (\ref{ring}) replacing $V^{(S)}(q)$ by 
$W_1$ + 2$W_2 k_F^2$ i.e.
\begin{equation} \label{VS1}
V_0^{(S=1)}= 
- s_0 - \frac{1}{6} s_3 \rho^{\gamma}
+\frac{1}{2} (s_2-s_1) k_F^2 
-\frac{1}{4} (s_1+s_2) q^2, 
\end{equation}
in the spin one case, and by
\begin{equation} \label{VS0}
V_0^{(S=0)}= 
s_0 +\frac{(\gamma+1)(\gamma+2)}{12} s_3 \rho^{\gamma}
+\frac{1}{2} (s_1+3 s_2) k_F^2 
+\frac{1}{2} (s_1- 3 s_2) q^2, 
\end{equation}
in the spin zero case.
From the expression for the structure function
\begin{equation} \label{strength}
S(\omega, {\bf q}, T )=-\frac{1}{\pi} \frac{ 2 {\Im}m \, 
\chi_0(\omega, {\bf q})} 
{(1-V_0^{(S=1)} {\Re}e \, \chi_0)^2+(V_0^{(S=1)} {\Im}m \,  \chi_0)^2} 
\frac{1}{1-e^{-\omega/T}} ,
\end{equation}
we see that, for a given value of the momentum ${\bf q}$,
zero sound will be present whenever there exists a solution 
$\omega_R$ of the equation
\begin{equation} \label{poles}
1-V_0^{(S)} {\Re}e \, \chi_0(\omega_R, {\bf q})=0,
\end{equation}
in the energy domain where the imaginary part of the bare
response function is small (or zero). At zero temperature this
domain corresponds to (see equation(\ref{IMCHI}))
\begin{equation} \label{omegap}
\omega_R \ge \omega_+= \frac{\hbar^2 q}{2m}(2k_F+q).
\end{equation}
At this point the real part of the response function is given by
equation(\ref{RECHI}) i.e.
\begin{equation}
{\Re}e \, \chi_0(\omega_+)= \frac{m^*}{4 \pi^2 \hbar^2} 
\{ (2k_F+q) \log (\frac{2k_F+q}{q}) - 2 k_F \}.
\end{equation}
Since the real part of the response function decreases with
energy in the domain of interest we find that the condition for
zero sound at zero temperature is
\begin{equation} \label{condition}
V_0^{(S)} {\Re}e \, \chi_0(\omega_+, {\bf q}) \ge 1.
\end{equation} 
For interaction SIII one finds that for a momentum q of the
order of 5 MeV/c and a density of 0.1 neutron per fm$^3$ one has
\begin{equation} \label{skyrme3}
V_0^{(S=1)}= 112.9 {\rm MeV} \times {\rm fm}^3~~~~{\rm and}~~~~
V_0^{(S=0)}=-560.8 {\rm MeV} \times {\rm fm}^3
\end{equation}
and
\begin{equation} \label{skyrme3s0}
{\Re}e \, \chi_0(\omega_+, {\bf q}) =6.2 \times 10^{-3}
{\rm MeV}^{-1} \times {\rm fm}^{-3}.
\end{equation} 
These values show that the residual particle-hole 
interaction in the S=0 channel has the wrong sign
to produce a zero sound mode. In contrast 
in the S=1 case it is repulsive and strong enough to give a zero 
sound mode. It is however not as strong as in the case of 
liquid helium-3 \cite{walecka} where zero sound occurs far 
in the quadratically decreasing tail of the real 
part of the response function. 
Here in contrast, the energy of the zero
sound mode is close to the maximum of the
real part of the response function, $\omega_{max}$= $q
k_F/m$. Note that this value belongs to the integration domain of
equation (\ref{opa}), namely
\begin{equation}
| \omega | \le q  \le 2 E_{\nu} - \omega~~~~~\omega \le E_{\nu},
\end{equation}
thus allowing, in principle, the zero-sound mode to contribute 
to the neutrino mean-free path. 
It is worthwhile noting that for a strong
particle-hole interaction the magnetic strength would be driven outside the
integration domain which would significantly increase the mean-free path.

\subsection{Energy density near the magnetic transition}

At the critical density the symmetric Hartree-Fock ground state
becomes unstable. This can be visualized looking at the
expression of the energy density of spin polarized neutron
matter:
\begin{equation} \label{EDENS}
{\cal H}= \frac{\hbar^2}{2m^*}\,
(\tau_{ \uparrow}+\tau_{ \downarrow})  +
(s_0+ \frac{1}{6}\,s_3\, \rho^{\gamma}) \,\rho_{ \uparrow}\rho_{
\downarrow}  
+
\frac{1}{8}(-s_1+s_2) (\rho_{\uparrow}-\rho_{ \downarrow})
(\tau_{ \uparrow}-\tau_{ \downarrow}).
\end{equation}
from which one can construct the  
energy per neutron as a function of the spin polarization parameter
\begin{equation} \label{XASYM}
x=\frac{\rho_{ \uparrow}-\rho_{ \downarrow}}{\rho}.
\end{equation}
The minimum energy per particle  occurs at $x$=0 for densities
smaller than the critical one and at nonvanishing asymmetries
for larger densities. It actually reaches rapidly the value
$x$=1 which corresponds to a fully polarized medium.

In the vicinity of the magnetic transition, the response
function in the S=1 channel behaves, according to equation (\ref{polar}),
as the inverse of the spin asymmetry coefficient
$a_{\sigma}$
in the limit $\omega$=0 and $q$=0. 
This can be checked by noting that 
in the limit of a small momentum transfer one has
\begin{equation} \label{RECHI0}
\Re e \chi_{0}(\omega,q=0, k_F,T=0)= - \frac{m^* k_F}{\pi^2 \hbar^2},
\end{equation}
and a negligible value of the imaginary part of the bare response function.
The RPA response thus becomes \cite{braghin}
\begin{equation}
\chi^{(S=1)}(\omega=0, {\bf q}=0)= - \frac{\rho}{2a_{\sigma}}.
\end{equation}
This formula accounts for the presence of the peak 
which develops near zero energy in
figure 2 when the density becomes close to its critical value.
It also shows that in the limit of a vanishing momentum transfer 
the mean-free path would exactly
vanish at the transition point.  
This result is not in conflict with the sum rule mentionned 
previously (see eq. (\ref{sumrule})) because a peak at zero energy 
gives no contribution to the energy weighted sum.
We have not attempted to push
our RPA calculations beyond the critical point, since more work
is needed to produce a reliable Skyrme type parametrization in
this region.

\subsection{Momentum and temperature dependence}

In Fig. 3, the mean free paths obtained with the interaction
SLy230b are displayed as functions of the density for six
combinations of $k_i$ and $T$.  It is worthwhile noticing that
the effect of increasing momentum and/or temperature is similar
and  mostly consists of an overall reduction of the scale.
Scaling properties were already discussed by Haensel and Jerzak
\cite{haensel}. As a result of their approximation scheme these
authors found that scaling the neutrino energy and the
temperature by a factor $\alpha$ produces a scaling of the mean
free path by a factor 1/$\alpha^3$. Comparing the curves $k_i
c=T$= 5 MeV and $k_i c=T$= 10 MeV we see that this scaling law
holds to a reasonable accuracy in our RPA calculations, although
the curvature of the uppermost line is somewhat too high.

\subsection{Effect of residual interactions}

An important issue regarding neutrino propagation is the effect
of the residual neutron-neutron interaction. Various answers to
this question are found in the literature. Indeed while Iwamoto
and Pethick \cite{iwamoto} conclude that Fermi liquid effects
increase the mean free path to some extent, Haensel and Jerzak
\cite{haensel} find that near normal density, the overall effect
of the nucleon -nucleon interaction on the mean free path is
rather weak.  In contrast, the recent paper by Reddy {\it et.
al.} \cite{pons} reports that many-body effects suggest
considerable reductions in the opacities compared to the free
gas estimates.  

Our results are  displayed for interactions SIII
and SLy230b in the cases $T=k_i c=5$ MeV and $T=k_i c=10$ MeV in
Fig. 4.  This figure shows the relative opacities
$\lambda^{-1}/\lambda^{-1}_{free}(m^*)$ where
$\lambda^{-1}_{free}(m^*)$ corresponds to a gas of free neutrons
with effective mass $m^*$, rather than $m$, i.e. the scale is
furnished by the opacity in the mean field.  In the vicinity of
the saturation density of nuclear matter it can be seen that
this relative opacity is close to unity.  This means that for
such densities, the major role of the interaction is to dress
the bare mass giving rise to the mean field density dependent
effective one $m^*$. If the reference for the relative opacity
is however chosen as $\lambda_{free}(m)$, then one finds that
interactions do make the medium more transparent.  Indeed at
normal density one has $m^*/m$= 0.695 for interaction SLy230b
and the smaller the mass, the smaller the  opacity.  In
particular, at zero temperature
\begin{equation}
\frac{1}{\lambda_{free}}=\frac{G_F^2 k_F^2}{24 \pi^3}
\left( \frac{m^* k_i}{m^*+k_F} \right)^3.
\end{equation}
Since the effective mass becomes small at high density, large
reductions can occur.  Nevertheless, at large enough densities
the dominant effect in Fig. 4 is the reduction of the mean free
due to the vicinity of the magnetic instability.

The reason for having RPA values of the mean-free path very
close to the mean field ones (at least in the region just above
nuclear density), in spite of the presence of a zero sound mode
at zero temperature, is that this mode is rapidly smeared as
temperature increases because of the strong Landau damping,
clearly visible on Fig. 2. Moreover when temperature rises there
is a strong reduction of the real part of the response function,
which may correspond to a factor 2-3 for a temperature of the
order of 5 MeV (see e.g. figure 1 of reference \cite{fabio}).  As
a result one finds that the term $V_0^{(S=1)} \chi$ in the
expression of the RPA response function is substantially smaller
than unity, which implies that RPA and mean field response
functions are nearly identical in the  density temperature range
considered here.

\section{Discussion}

In the present work we have calculated neutrino propagation in
neutron matter within  a microscopic and consistent framework
able to describe both the nuclear equation of state and
dynamical structure factors.  The approach adopted here is based
on Skyrme type effective interactions and linear response (RPA)
theory. As already mentioned, an advantage of the present
procedure is the treatment of single particle and collective
contributions to the structure factors on an equal footing at
any temperature,  making unnecessary to postulate a
factorization of the temperature dependence of the structure
factors.  Such factorizations, which are in conflict with the
fact that spin zero sound disappears at temperatures of a few
MeV \cite{hnp}, overestimate the contribution of collective
modes.  Our calculation indicates that magnetic zero sound modes
appear at zero temperature for all Skyrme parametrizations here
considered, even at the largest momenta (i.e., q = 100 MeV/c) up
to density $\rho_c$; however, at finite temperatures they become
subject to Landau damping on the ph continuum, as shown in Fig.
2 and discussed in section III.

In the domain where the RPA is reliable (stable symmetric ground
state in the mean field picture) we have found that the most
important effects  of the interaction are the renormalization of
the mass of the neutrons into their effective mass $m^*$ in the
response function.  This makes the medium more transparent, thus
giving rise to shorter deleptonization time scales \cite{pons}.

We have found that at high density instabilities occur, which
signal a breakdown of the Skyrme like parametrization.  The
possible existence of a magnetic transition in neutron matter,
as well as its origin, have been investigated by several authors
\cite{rice,ostgaard,pandha,vidaurre,niembro,marcos,uma}. 
In general, one should expect that the
short-range repulsion of the nuclear interaction  originates
spin alignment at sufficiently high densities; this has been
verified in various model calculations  extending from
relativistic approaches\cite{niembro,marcos} to HF description
with finite range effective interactions\cite{uma}.  The effect
of a short range repulsion in neutron matter is most
conveniently illustrated in the case of a Skyrme force without
velocity dependent terms.  In this case a transition takes place
provided the density dependent repulsion grows rapidly enough.
Indeed, a zero range interaction does not contribute to the
potential energy in fully polarized neutron matter, because the
Pauli principle forbids configurations of two neutrons with  the
same spin at the same point. As a result, when the repulsion in
spin symmetric neutron matter exceeds $(2^{1/3}-1)$ times the
kinetic energy, the cost in kinetic energy requested to turn all
spins up becomes balanced by the gain in potential energy, thus
making the fully polarized state more favorable. The previous
analysis also shows how to work out interactions without
magnetic transitions. One needs a weak density dependent
repulsion,  more realistic to reproduce the monopole resonance,
in order to reduce the gain in potential energy. To explore the
results which would be obtained with such a force we have
slightly modified the interaction SLy230b so as to increase the
value of the critical density up to a factor 2, without
impairing the equation of state. The results are indicated in
Fig. 3 as SLy* for $T=5$ MeV (solid line), $T=10$ MeV (dashed
line), and $k_i c= 5$ MeV in both cases.  We find that the mean
free path remains roughly constant up to  densities around
0.9fm$^{-3}$.

Further studies are obviously needed in the vicinity of the
magnetic transition. Indeed, it occurs in a region where inputs
from more realistic calculations are necessary for the
determination of a reliable Skyrme type parametrization.
Furthermore, other types of phase transitions, such as
strangeness production or a quark-hadron transition, are likely
to occur before one reaches this region.

\acknowledgements 
One of us (D.V.) wishes to thank inspiring discussions with J.L.
Basdevant and  Ph. Chomaz which originated this paper. Helpful
comments from A.  P\'erez-Canyellas and J. Pons are gratefully
acknowledged. This work has been partly supported by grants
PB97-1139 (Spain) and PICT 0155/97 (Argentina).

\newpage

\begin{figure}
\vspace{10pc}
\centerline{\epsfysize=6.in \epsfbox{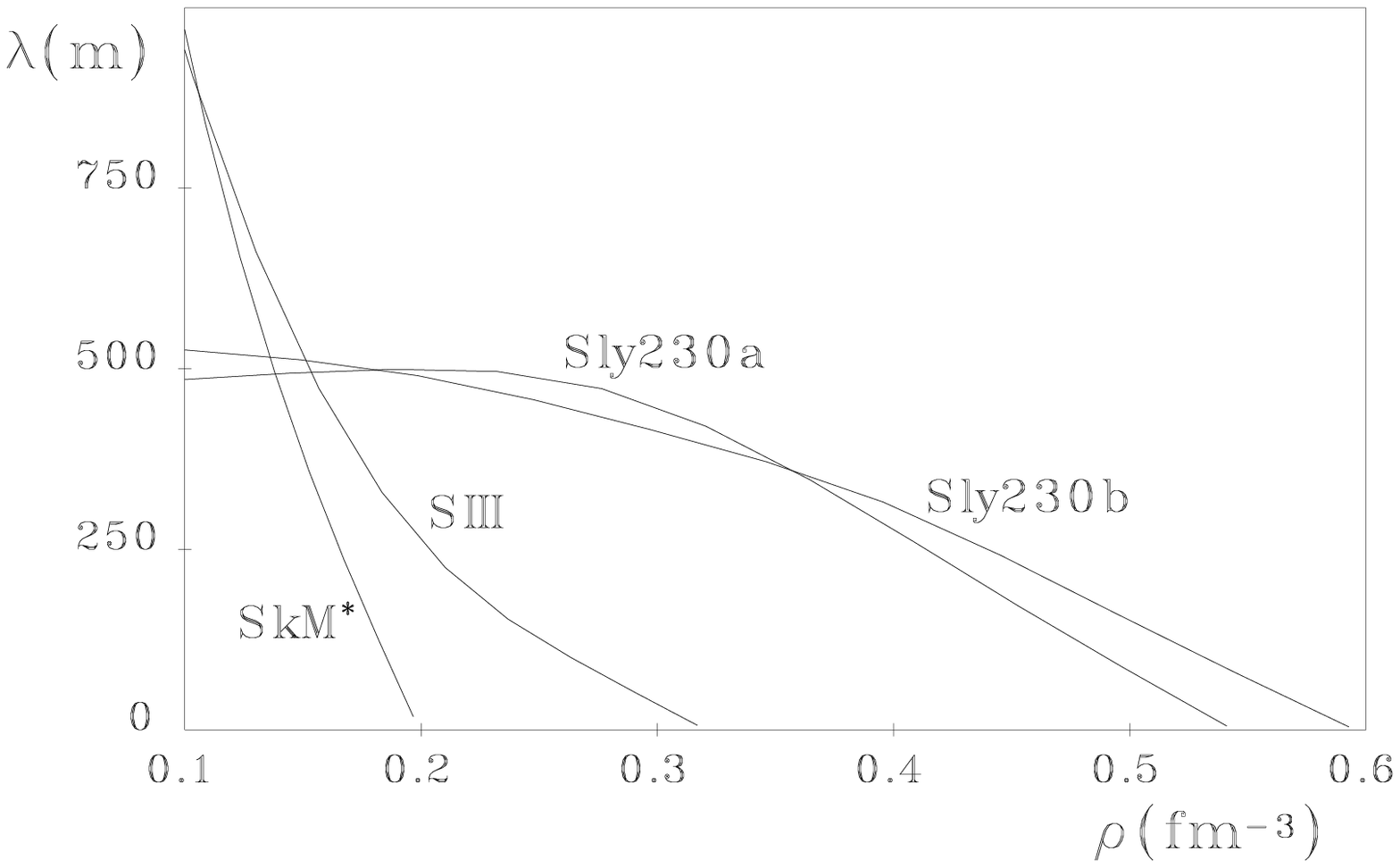}  }
\vspace{1pc}
\caption{The scattering mean free path for the neutrino as a
function of neutron density, for various Skyrme force
parametrizations.}
\end{figure}

\newpage

\begin{figure}
\centerline{\epsfysize=5.in \epsfbox{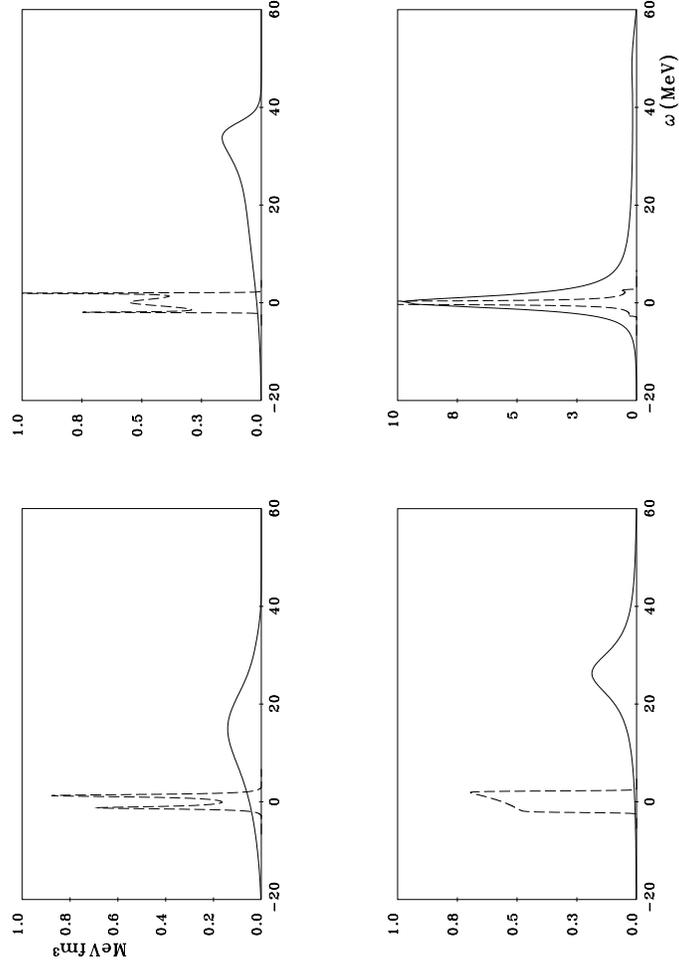}  }
\vspace{1pc}
\caption{Dynamical structure factors S$^{(S)}(\omega,q)$ for S=0
and 1 channels (left and right parts, respectively),  
as functions of energy  for interaction
SIII at a temperature $T$ =  5 MeV. Upper and lower parts 
respectively correspond to densities  0.1 fm$^{-3}$ and $\rho_c$.
Dashed and solid lines respectively correspond to transferred
momenta $q$= 5 and 100 MeV/c.}
\end{figure}

\newpage

\begin{figure}
\centerline{\epsfysize=4.in \epsfbox{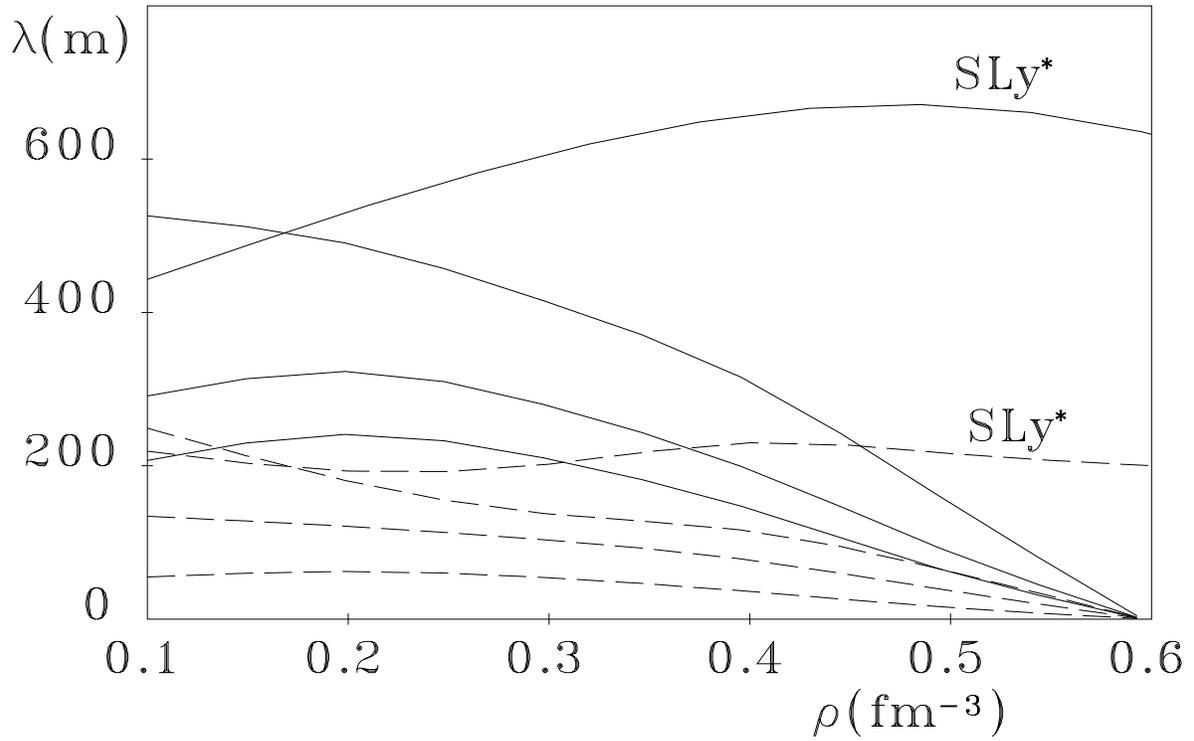}  }
\vspace{1pc}
\caption{The scattering mean free path for the neutrino
as a function of neutron density for the parametrization
SLy230b and its modificated version SLy$^*$.  Full and dashed
lines respectively correspond to 
temperatures 5 and 10 Mev. In each group, from above to below
the lines correspond to incoming momenta $k_i$ = 5, 10 and 3 T
(in MeV/c). 
}
\end{figure}

\newpage
\begin{figure}
\centerline{\epsfysize=6.in \epsfbox{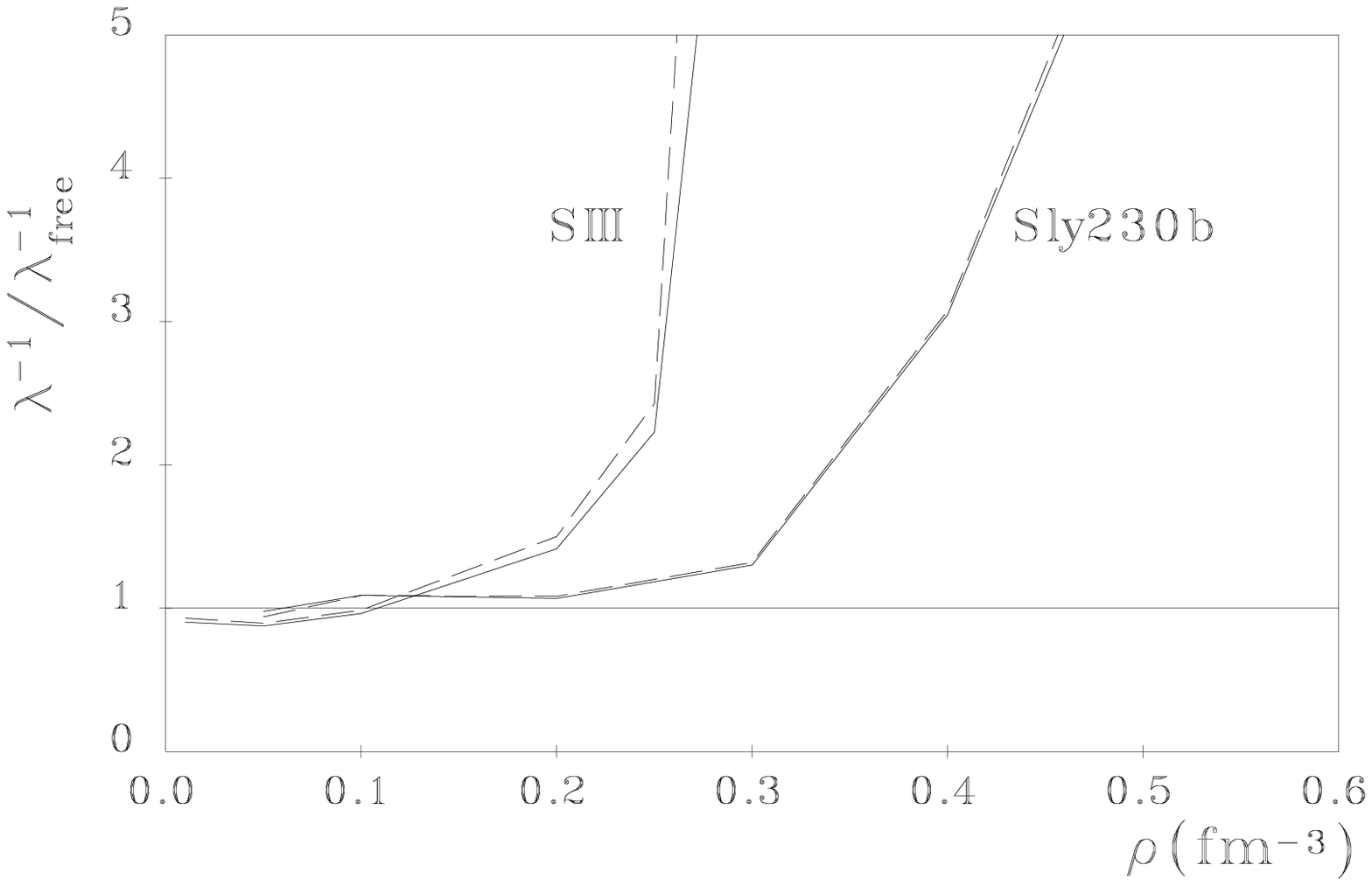}  }
\vspace{1pc}
\caption{The relative opacity $\lambda^{-1}/
\lambda^{-1}_{free}$ as a function of the density for interactions
SIII and SLy230b. Full and dashed lines
respectively correspond to $T = k_i c$ = 5 and 10 MeV. A
horizontal line corresponding to relative opacity equal to unity
has been drawn for easier comparison.}
\end{figure}

\end{document}